\documentclass[useAMS]{mn2e}
\usepackage{graphicx}
\usepackage{txfonts}
\usepackage{color}

\title[A high-frequency study of the Sunyaev-Zel'dovich effect morphology in galaxy clusters]{A high-frequency study of the Sunyaev-Zel'dovich effect morphology in galaxy clusters}

\author[D. A. Prokhorov et al.]{D. A. Prokhorov$^{1}$\thanks{E-mail:
phdmitry@stanford.edu}, S. Colafrancesco$^{2}$, T. Akahori$^{3}$, E.
T. Million$^4$, S. Nagataki$^{5}$,
\newauthor
K. Yoshikawa$^{6}$\\
$^{1}$Hansen Experimental Physics Laboratory, Stanford University,
Stanford, CA 94305, USA\\
$^{2}$ INAF - Osservatorio Astronomico di Roma via Frascati 33, I-00040 Monteporzio, Italy. Email: sergio.colafrancesco@oa-roma.inaf.it\\
$^{3}$ Research Institute of Basic Science, Chungnam National University, Daejeon, Republic of Korea\\
$^{4}$ University of Alabama, Department of Physics and Astronomy,
206 Gallalee Hall Box 870324, Tuscaloosa AL 35487\\
$^{5}$ Yukawa Institute for Theoretical Physics, Kyoto University,
Kitashirakawa Oiwake-cho, Sakyo-ku, Kyoto, 606-8502, Japan\\
$^{6}$ Center for Computational Sciences, University of Tsukuba,
1-1-1, Tennodai, Ibaraki 305-8577, Japan}

\begin{document}

\pagerange{\pageref{firstpage}--\pageref{lastpage}} \pubyear{2002}

\maketitle

\label{firstpage}

\begin{abstract}

High-frequency, high-resolution imaging of the Sunyaev-Zel'dovich
(SZ) effect is an important technique to study the complex
structures of the atmospheres of merging galaxy clusters. Such
observations are sensitive to the details of the electron spectrum.
We show that the morphology of the SZ intensity maps in simulated
galaxy clusters observed at 345 GHz, 600 GHz, and 857 GHz are
significantly different because of SZ relativistic corrections.
These differences can be revealed by high-resolution imaging
instruments.

We calculate relativistically corrected SZ intensity maps of a
simulated, massive, merging galaxy cluster and of the massive,
merging clusters 1E0657-558 (the Bullet Cluster) and Abell 2219. The
morphologies of the SZ intensity maps are remarkably different
between 345 GHz and 857 GHz for each merging cluster. We show that
high-resolution imaging observations of the SZ intensity maps at
these frequencies, obtainable with the LABOCA and HERSCHEL-SPIRE
instruments, allow to fully exploit the astrophysical relevance of
the predicted SZ morphological effect.
\end{abstract}


\section{Introduction}

Galaxy clusters are the largest gravitationally bound structures in
the Universe, with sizes of the order Mpcs. The space between
galaxies in the most massive clusters is filled with a low-density
($\sim 10^{-2}$--$10^{-3}$ cm$^{-3}$) high temperature
($k_{\mathrm{B}} T\sim 10-15$ keV), highly ionized X-ray emitting
plasma. Inverse Compton scattering of hot free electrons in clusters
of galaxies by the cosmic microwave background (CMB) radiation field
causes a change in the spectrum of the CMB radiation towards
clusters of galaxies (the Sunyaev-Zel'dovich effect, hereinafter the
SZ effect; see Sunyaev \& Zel'dovich 1980). The SZ effect measures
the pressure of the electron population integrated along the line of
sight. Relativistic effects are significant for high temperature
plasmas in massive galaxy clusters (see, e.g., Rephaeli 1995). A
relativistically correct formalism for the SZ effect that is based
on the probability distribution of the photon frequency shift after
scattering was given by Wright (1979) to describe the Comptonization
process of soft photons by high temperature plasma. We use the
Wright formalism in this paper, which has been proven to be
relativistically covariant (Nozawa \& Kohyama. 2009, Colafrancesco
\& Marchegiani 2010). The relativistic treatment of the SZ effect
allows us to measure the temperature of intracluster plasma (see
Pointecouteau et al. 1998; Hansen et al. 2002) and even the cluster
temperature profile (Colafrancesco \& Marchegiani 2010).
Relativistic corrections of the SZ effect from massive galaxy
clusters are significant at high frequencies (above $\sim 300$ GHz).
The first detection of the SZ effect increment at frequencies above
600 GHz, where the relativistic corrections of the SZ effect
dominate, has recently been obtained with HERSCHEL-SPIRE (Zemcov et
al. 2010) for the cluster 1E0657-558, the so-called Bullet Cluster.
The present-day and planned microwave and millimeter instruments are
able to provide spatially resolved observations of the SZ intensity
maps for extended galaxy clusters, approaching the precision of
X-ray observations. In fact, the LABOCA bolometer camera on APEX
provides us with SZ intensity maps with an angular resolution of
19.5$^{\prime\prime}$ at a frequency of 345 GHz (e.g. Nord et al.
2010), while HERSHEL-SPIRE allows to obtain SZ intensity maps with
angular resolutions of 36 $^{\prime\prime}$ at 600 GHz and
25$^{\prime\prime}$ at 857 GHz (Zemcov et al. 2010). Such
instruments are promising to reveal the morphology of the SZ
intensity maps in order to study the complex structures of the
atmospheres of massive merging clusters.

In this paper we use the results of both a simulated, hot, merging
cluster and X-ray observations of the clusters 1E0657-558 and A2219
to show that the SZ relativistic effect provides us with a method to
study the complex structure of the electron distribution in massive
merging clusters. We calculate SZ effect intensity maps at 345 GHz,
600 GHz and 857 GHz and show that the difference in their
morphologies could be revealed by LABOCA and HERSCHEL-SPIRE.

Both SZ measurements and X-ray observations allow us to determine
the gas temperature of hot merging clusters of galaxies. We will
show that an analysis of the morphology of SZ intensity maps at high
frequencies provides us with a method to study high-temperature
regions in galaxy clusters. This method is more suitable to analyze
high-temperature plasmas compared with the method based on X-ray
observations because of uncertainties in the determination of the
gas temperature from X-ray measurements. Note that the X-ray surface
brightness is proportional to the factor of
$\exp(-E/k_{\mathrm{b}}T_{\mathrm{e}})$ and that this dependence
provides us with a possibility to derive the gas temperature, where
$E$ is the X-ray photon energy, $T_{\mathrm{e}}$ is the electron
temperature and $k_{\mathrm{b}}$ is the Boltzmann constant. However,
this exponential factor slightly depends on the gas temperature if
the gas temperature is significantly higher than the energy of an
emitted X-ray photon. The modern X-ray instruments (Suzaku, Chandra,
and XMM-Newton) have the energy range of (0.5 keV-10 keV) and a
small effective area at energies higher 8 keV. Therefore, the gas
temperature of hot regions of galaxy clusters is derived with a
large uncertainty from X-ray observations. Furthermore, X-ray
spectra of high-temperature merging clusters can be contaminated by
a non-thermal-like X-ray component (see Million \& Allen 2009) which
is possibly caused by the Inverse Compton effect of CMB photons on
ultra-relativistic electrons.

The layout of the paper is as follows. We describe the morphological
properties of the SZ effect in Sect. 2. We calculate the SZ
intensity maps at 345 GHz, 600 GHz, and 857 GHz (in the framework of
the Wright formalism) for the simulated galaxy cluster and show in
Sect. 3.1 that the evidence for the morphological SZ effect from
this cluster could be revealed by the combined analysis of LABOCA
and HERSCHEL-SPIRE. The morphological properties of the SZ effect
from the clusters 1E0657-558 and A2219 are considered in Sects. 3.2
and 3.3, respectively. In Sect. 3.4, we demonstrate that the ratio
of the SZ intensities at frequencies of 600 GHz to 345 GHz can be
used to quantify the difference in the morphology of the SZ
intensity maps. We show that the formalism based on the generalized
Kompaneets equation can be applied for calculating the SZ intensity
maps in Sect. 3.5, and our conclusions are presented in Sect. 4.

\section{Morphology of the SZ effect in massive galaxy clusters}

The CMB intensity change produced by the SZ effect of
non-relativistic electrons in the framework of the Kompaneets
approximation is (for a review, see Birkinshaw 1999)
\begin{equation}
\Delta I_{nr}(x) = I_{\mathrm{0}} g(x) y_{\mathrm{gas}}, \label{Inr}
\end{equation}
where $x=h\nu/k_{\mathrm{b}} T_{\mathrm{cmb}}$, $I_{\mathrm{0}}=2
(k_{\mathrm{b}} T_{\mathrm{cmb}})^3 / (hc)^2$,
$T_{\mathrm{cmb}}=2.725$ K, and the spectral function $g(x)$ is
given by
\begin{equation}
g(x)=\frac{x^4 \exp(x)}{(\exp(x)-1)^2} \left(x
\frac{\exp(x)+1}{\exp(x)-1}-4\right).
\end{equation}
The subscript  $`nr' $ denotes that Eq. (\ref{Inr}) was obtained in
the non-relativistic limit. The Comptonization parameter
$y_{\mathrm{gas}}$ is given by
\begin{equation}
y_{\mathrm{gas}}=\frac{\sigma_{\mathrm{T}}}{m_{\mathrm{e}}c^2} \int
n_{\mathrm{gas}} k T_{\mathrm{e}} dl,
\end{equation}
where the line-of-sight integral extends from the last scattering
surface of the CMB radiation to the observer at redshift z=0,
$T_{\mathrm{e}}$ is the electron temperature, $n_{\mathrm{gas}}$ is
the number density of the gas, $\sigma_{\mathrm{T}}$ is the Thomson
cross-section, $m_{\mathrm{e}}$ the electron mass, $c$ the speed of
light, $k_{\mathrm{b}}$ the Boltzmann constant, and $h$ the Planck
constant.

Note that all the SZ intensity maps derived in the framework of the
Kompaneets approximation have the same spatial morphology at each
frequency, since the spectral function $g(x)$ does not depend on gas
temperature. The morphologies of the SZ intensity simulated maps at
frequencies $\nu_1 = 128$ GHz and $\nu_2 = 369$ GHz are similar for
the cool merging cluster in Prokhorov et al. (2010b). This is
because the SZ effect from cool galaxy clusters can approximately be
described in the framework of the non-relativistic Kompaneets
approximation.
However, we will show here that the relativistic corrections of the
SZ effect can significantly change the morphology of the SZ
intensity maps especially at high frequencies.

The CMB intensity change in the Wright formalism can be written in
the form proposed by Prokhorov et al. (2010a) given by
\begin{equation}
\Delta I(x) = I_{\mathrm{0}}
\frac{\sigma_{\mathrm{T}}}{m_{\mathrm{e}}c^2} \int n_{\mathrm{gas}}
k_{\mathrm{b}} T_{\mathrm{e}} G(x, T_{\mathrm{e}}) dl, \label{form}
\end{equation}
where the spectral function $g(x)$ is replaced by the relativistic
spectral function $G(x, T_{\mathrm{e}})$, which depends explicitly
on the electron temperature. The relativistic spectral function
$G(x, T_{\mathrm{e}})$ derived in the framework of the Wright
formalism is given by
\begin{equation}
G(x, T_{\mathrm{e}})=\int^{\infty}_{-\infty} \frac{P_{1}(s,
T_{\mathrm{e}})}{\Theta(T_{\mathrm{e}})} \left(\frac{x^3 \exp(-3
s)}{\exp(x \exp(-s))-1}-\frac{x^3}{\exp(x)-1}\right) ds, \label{G}
\end{equation}
where $\Theta(T_{\mathrm{e}}) = k_{\mathrm{b}} T_{\mathrm{e}}/
m_{\mathrm{e}}c^2$, and $P_{1}(s, T_{\mathrm{e}})$ is the
distribution of frequency shifts for single scattering (Wright 1979;
Rephaeli 1995; for a review, see Birkinshaw 1999), and $s$ is the
logarithmic frequency shift caused by a scattering (see Birkinshaw
1999).

\begin{figure}
\centering
\includegraphics[angle=0, width=8.0cm]{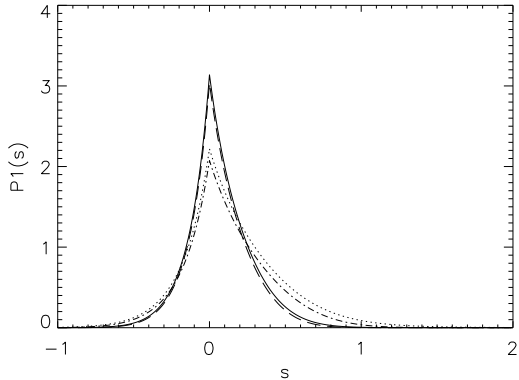}
\caption{The solid and dotted lines show the approximate scattering
kernels calculated by using the Sunyaev (1980) approach for thermal
plasmas at 10 keV and 20 keV, respectively. The dashed and
dash-dotted lines show the scattering kernels calculated according
to Eq. (33) from Birkinshaw (1999) for thermal plasmas at 10 keV and
20 keV, respectively.} \label{Fig1}
\end{figure}

We calculated the scattering kernels (the distribution of frequency
shifts for single scattering), $P_1(s, T_{\mathrm{e}})$, for thermal
plasmas at 10 keV and 20 keV by using Eq. (33) of Birkinshaw (1999),
which is a relativistically correct expression. These scattering
kernels are shown in Fig. \ref{Fig1} by dashed and dash-dotted lines
for thermal plasmas at 10 keV and 20 keV, respectively. We also
calculated the approximate scattering kernels for thermal plasmas at
10 keV and 20 keV by using Eq. (4) of Sunyaev (1980), which is based
on the expression of the source term (see Appendix I of
Babuel-Payrissac \& Rouvillois 1970) calculated in the
non-relativistic limit. The approximate scattering kernels are shown
in Fig. \ref{Fig1} by solid and dotted lines for thermal plasmas at
10 keV and 20 keV, respectively. We find that the approximate
scattering kernels tend to have a stronger tail at large values of
s, this agrees with the comparison of the approximate and
relativistically correct kernels performed by Birkinshaw (1999). The
non-relativistic expression of the source term derived by
Babuel-Payrissac \& Rouvillois (1970) was generalized by Sazonov \&
Sunyaev (2000), who included terms describing relativistic effects
in the source term and analytically derived an approximate
expression for the source term which is valid for a high-temperature
plasma. Since we apply a numerical method to calculate the SZ effect
in this paper, we use the relativistically correct expression for
the scattering kernel described by Birkinshaw (1999). Below we
compare the relativistically corrected SZ signal with that obtained
in the framework of the Kompaneets formalism (which is derived in
the diffusion approximation) because the Kompaneets formalism is
usually used to describe the non-relativistic SZ effect.

Since the relativistic spectral function depends on the electron
temperature, the SZ intensity maps of massive merging galaxy
clusters should appear different at higher and higher frequencies.
This is due to the increasing dominance of SZ relativistic
corrections and to the fact that merging clusters are expected to
produce complex temperature distributions. Evidence for the
relativistic effects on the SZ intensity maps will be called here,
for the sake of conciseness, as the 'morphological SZ effect'. To
study the properties of the morphological SZ effect we calculate the
SZ intensity maps from a hydrodynamic simulation of a massive
merging cluster presented in Akahori \& Yoshikawa (2010) and for the
clusters 1E0657-558 and A2219 analyzed by Million \& Allen (2009).
We expect two morphological SZ effects: i) an increase of the SZ
intensity at very high frequencies caused by the presence of shocked
regions, and ii) a decrease of the SZ intensity at very high
frequencies caused by the presence of cool substructures.

We have checked that the contribution of the kinematical SZ effect
at very high frequencies (such as 857 GHz) is negligible compared
with that of the SZ relativistic corrections for high temperature
plasmas. Therefore, SZ observations at very high frequencies are an
important tool to reveal unusually hot and cold regions in massive
merging clusters of galaxies.

\section{Morphology simulations}

We calculate the SZ intensity maps for the simulated cluster in
Sect. 3.1 and for the 1E0657-558 and A2219 galaxy clusters in Sects.
3.2 and 3.3, respectively. We study the ratio of the SZ intensity
signals at two frequencies in Sect. 3.4 and show that the
generalized Kompaneets formalism can be used instead of the Wright
formalism to derive the SZ intensity maps in Sect. 3.5. We now
discuss SZ measurements which can be performed with the LABOCA and
HERSCHEL-SPIRE instruments.

Note that the intensity of $I_{\mathrm{0}}=2 (k_{\mathrm{b}}
T_{\mathrm{cmb}})^3 / (hc)^2$ equals $6.0 \rm{mJy/arcsec^2}$ and
that the SZ intensity, $\Delta I$, does not depend on redshift of
galaxy clusters (see, e.g., Birkinshaw 1999).

In Sects. 3.1, 3.2, and 3.3, we calculate the SZ effect from a
simulated massive galaxy cluster and from the 1E0657-558 and A2219
galaxy clusters at frequencies of 345 GHz, 600 GHz, and 857 GHz and
demonstrate that the SZ intensities $\Delta I/I_{0}$ at these
frequencies are of the order of $1.5\times10^{-3}$,
$8.0\times10^{-4}$, and $2.0\times10^{-4}$, respectively. Using the
beam sizes of LABOCA at 345 GHz and of HERSCHEL-SPIRE at 600 GHz and
857 GHz, we derive the SZ intensities in the units of mJy/beam.

The SZ intensity $\Delta I/I_{0}$ of $1.5\times10^{-3}$ at a
frequency of 345 GHz corresponds to 3.5 mJy/beam. We find a noise
rms value of 1.4 mJy/beam in the LABOCA map, using the LABOCA
observing time calculator
\footnote{http://www.apex-telescope.org/bolometer/laboca/obscalc/}
for the integration time of 55 hrs. Therefore, the SZ signal to
noise ratio (which can be measured by LABOCA with the integration
time of 55 hrs) is 2.5. To measure the SZ signal of 3.5 mJy/beam
with the signal to noise ratio of 4, we find that the integration
time of 130 hrs is required.

The SZ intensity $\Delta I/I_{0}$ of $8.0\times10^{-4}$ at a
frequency of 600 GHz corresponds to 6.3 mJy/beam. The SZ intensity
$\Delta I/I_{0}$ of $2.0\times10^{-4}$ at a frequency of 857 GHz
corresponds to 0.75 mJy/beam. The instrument noise of HERSCHEL-SPIRE
(Nguyen et al. 2010) can be reduced with integration time. Using the
Herschel Observation Planning Tool
\footnote{http://herschel.esac.esa.int/Tools.shtml}, we find that
the total on-source integration time of 1 hr is sufficient to make
the instrumental noise level of 1.0 mJy/beam for a frequency of 600
GHz much smaller than the expected SZ signal. The total on-source
integration time of 4 hrs is sufficient to make the instrumental
noise level of 0.25 mJy/beam for a frequency of 857 GHz. Foreground
contamination is a source of additional noise at frequencies of 600
GHz and 857 GHz. The foreground contamination can be subtracted from
the 350 $\mu$m (corresponding to 857 GHz) and 500 $\mu$m
(corresponding to 600 GHz) channels by analyzing the 250 $\mu$m
emission. Zemcov et al. (2010) generated a 250 $\mu$m source catalog
and corrected the 350 $\mu$m and 500 $\mu$m  emission maps to obtain
foreground-free measurements of the SZ effect.

\subsection{The morphological SZ effect from a simulated merging
cluster}

\begin{figure}
\centering
\includegraphics[angle=0, width=7cm]{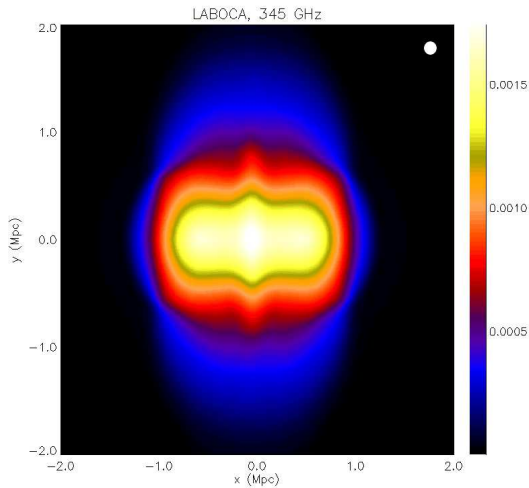}
\caption{The SZ intensity map $\Delta I/I_{0}$ for the simulated
cluster at a frequency of 345 GHz smoothed to the resolution of
LABOCA of 19$^{\prime\prime}$} \label{F1}
\end{figure}

\begin{figure}
\centering
\includegraphics[angle=0, width=7cm]{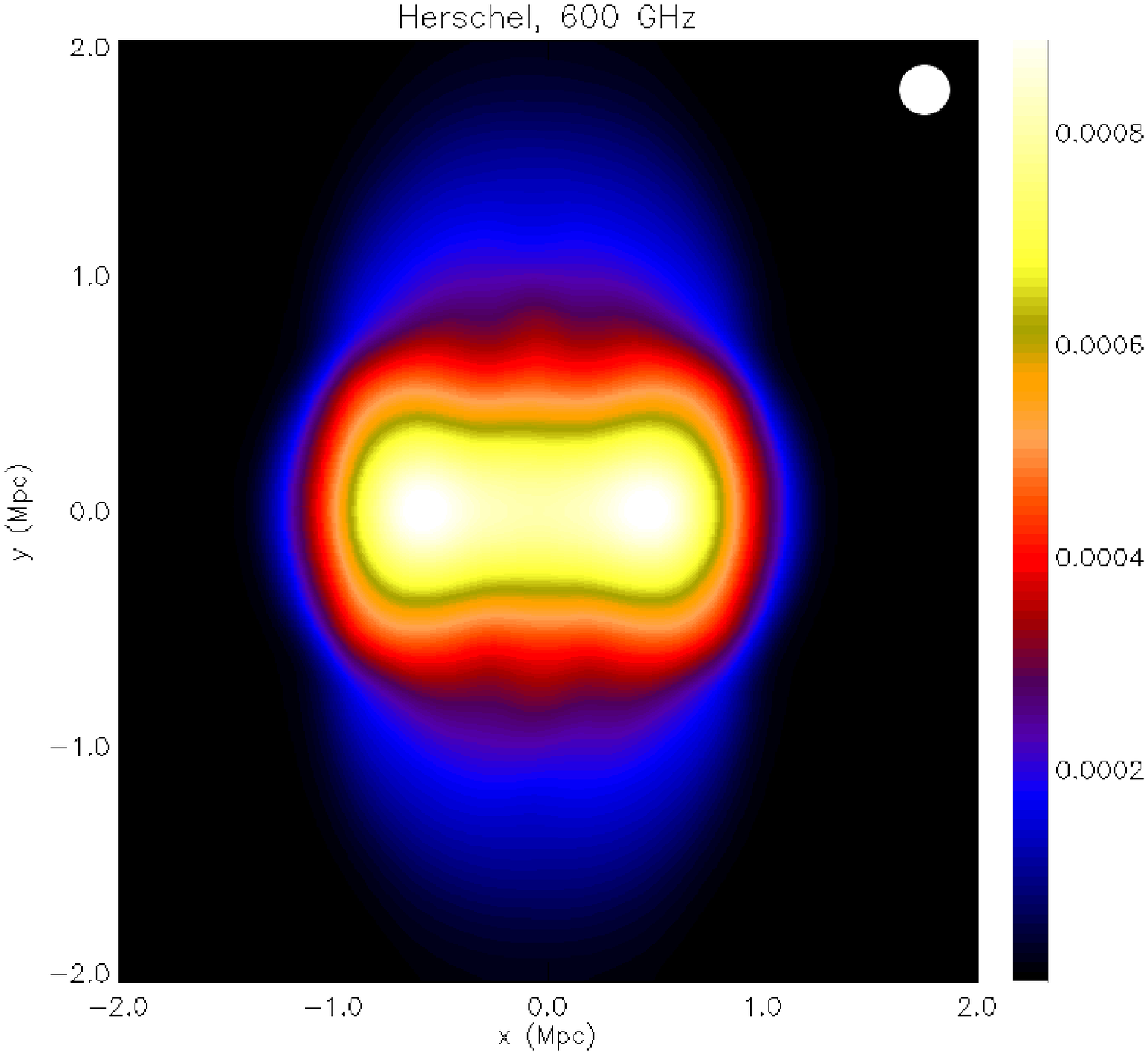}
\caption{The SZ intensity map $\Delta I/I_{0}$ for the simulated
cluster at 600 GHz smoothed to the resolution of HERSHEL-SPIRE of
36$^{\prime\prime}$}\label{F2}
\end{figure}

\begin{figure}
\centering
\includegraphics[angle=0, width=7cm]{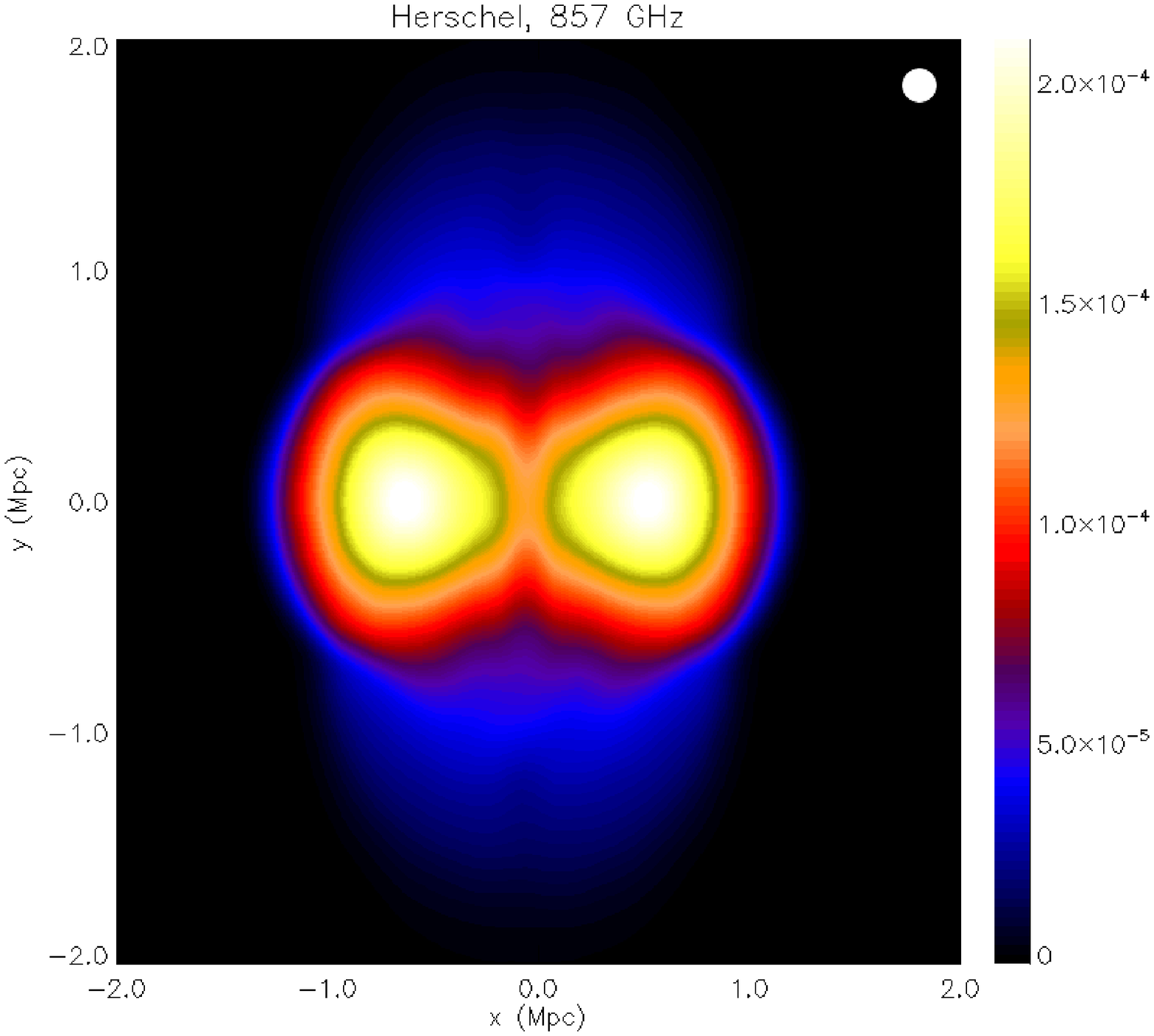}
\caption{The SZ intensity map $\Delta I/I_{0}$ for the simulated
cluster at 857 GHz smoothed to the resolution of HERSHEL-SPIRE of
25$^{\prime\prime}$}\label{F22}
\end{figure}

To show that the morphologies of the SZ intensity maps of a high
temperature merging galaxy cluster, derived in the Wright formalism,
are significantly different at 345 GHz, 600 GHz, and 857 GHz, we use
the 3D numerical simulations of the merging hot galaxy cluster
presented in Akahori \& Yoshikawa (2010). These authors considered
an encounter of two free-falling galaxy clusters from the epoch of
their turn-around (for a review, see Sarazin 2002). We assume in
this simulation that the two galaxy clusters have equal virial
masses of $M_{\mathrm{vir}}=8\times10^{14} M_\odot$, and that the
impact parameter is zero (for detail on the simulation, see Akahori
\& Yoshikawa 2010). 3D numerical simulations are an interesting tool
which provides us with precisely determined 3D gas temperature maps,
while X-ray data have large uncertainties in a determination of the
gas temperature in high-temperature regions of galaxy clusters.
Therefore, numerical simulations are a useful tool to test
theoretical predictions of the morphological SZ effect.

A possible candidate for a high temperature cluster undergoing a
major cluster merger is MACSJ0717.5+3745 at z=0.545 (Ma et al.
2009). The angular extent of the simulated box of 4 Mpc$\times$4 Mpc
at this redshift is close to $7^{\prime}\times7^{\prime}$ and is
covered by the field of views (FOVs) of LABOCA and HERSCHEL-SPIRE.
To obtain SZ effect intensity maps smoothed to the resolution of
LABOCA and HERSCHEL-SPIRE, we locate the simulated cluster at a
redshift z=0.545.

To produce the SZ intensity maps at $\nu=345$ GHz, 600 GHz, and
$857$ GHz we use the 3D density and temperature maps (see Figs. 2
and 3 from Akahori \& Yoshikawa 2010) for the simulated merging
galaxy cluster at a time of $t=0.5$ Gyr, where $t=0$ Gyr corresponds
to the time of the closest approach of the centers of the dark
matter halos. We finally calculated the SZ intensity maps using the
Wright formalism in the approach described in Prokhorov et al.
(2010a, 2010b). The SZ intensity map at a frequency of 345 GHz
derived from the simulated maps of the gas density and temperature,
and smoothed to the resolution of FWHM of LABOCA
(19$^{\prime\prime}$) is shown in Fig. \ref{F1}. \footnote{We show
the beam size of the corresponding instruments by a circle in the
upper corners of all the Figs. on which the SZ intensity maps are
plotted.} We calculate the SZ intensity maps at frequencies of 600
GHz and 857 GHz derived from the simulated maps of the gas density
and temperature, and smoothed to the resolution of FWHM of
HERSCHEL-SPIRE (36$^{\prime\prime}$ and 25$^{\prime\prime}$,
respectively). The calculated SZ intensity maps at frequencies of
600 GHz and 857 GHz are shown in Fig. \ref{F2} and \ref{F22},
respectively.

We find that the maximum of the SZ intensity increment at a
frequency of 345 GHz (see Fig. \ref{F1}) is at the center of this
map, while the maximal values of the SZ intensity increment at
frequencies of 600 GHz and 857 GHz (see Figs. \ref{F2} and
\ref{F22}) are in the post-shock regions (see the Mach number
distribution of ICM in Fig. 2 of Akahori \& Yoshikawa 2010). This is
because the gas temperature in the post-shock regions is the highest
and SZ relativistic corrections are significant in these regions.

Figures \ref{F1} and \ref{F22} show that this difference in the SZ
intensity maps at frequencies of 345 GHz and 857 GHz could be
revealed by the combined analysis of the LABOCA and HERSCHEL-SPIRE
SZ intensity maps. This is because the difference between the
calculated central SZ signal at 857 GHz and the expected central SZ
increment at 857 GHz, derived from the SZ intensity map at 345 GHz
by using the Kompaneets approximation, is $\Delta I/I_{0}\approx
1.0\times10^{-4}$, which corresponds to 0.38 mJy/beam. The on-source
integration time of foreground-free HERSCHEL-SPIRE observations of 2
hrs is sufficient to reduce the instrumental noise for a detection
of the morphological SZ effect at 857 GHz. Analyzing the SZ
intensity maps at frequencies of 345 GHz and 600 GHz (see Figs.
\ref{F1}, \ref{F2}, and also Fig. \ref{F7}, which shows the ratio of
the SZ signals at frequencies of 600 GHz to 345 GHz), we find that
foreground-free SZ measurements at 600 GHz by HERSCHEL-SPIRE with
the sensitivity of $\Delta I/I_{0}\approx 2.0\times10^{-4}$ (which
corresponds to 1.6 mJy/beam) are required to reveal the difference
in morphologies in the central SZ map region. Therefore, the
on-source integration time of 45 min is necessary to reduce the
instrumental noise for a detection of the difference in morphologies
between the calculated and expected (from the SZ intensity map at
345 GHz by using the Kompaneets approximation) SZ intensity maps at
600 GHz in the central region.

\subsection{The morphological SZ effect from the cluster 1E0657-558}

We calculate the SZ intensity maps in the Wright formalism for the
cluster 1E0657-558 (the Bullet Cluster) to show that the
morphological SZ effect from this cluster could be revealed by
observations of the SZ signal at frequencies of 345 GHz, 600 GHz and
857 GHz.

The cluster 1E0657-558 at z=0.296 consists of two colliding
subclusters: a less massive (with $kT=6$ keV) western subcluster and
a more massive (with $kT=14$ keV) eastern subcluster (e.g.
Markevitch et al. 2002).
The peak of the X-ray surface brightness is centered on the
fast-moving, merging western subcluster, i.e. the `Bullet', rather
than the overall X-ray center. We use the temperature and pressure
maps of the cluster 1E0657-558 which were obtained by Million \&
Allen (2009) by analyzing Chandra observations. These maps include
the most interesting morphological features of this cluster and have
the angular extent close to $6^{\prime}\times6^{\prime}$. The
analyzed region can be therefore covered by the FOVs of LABOCA and
HERSCHEL-SPIRE instruments.

\begin{figure}
\centering
\includegraphics[angle=0, width=7.0cm]{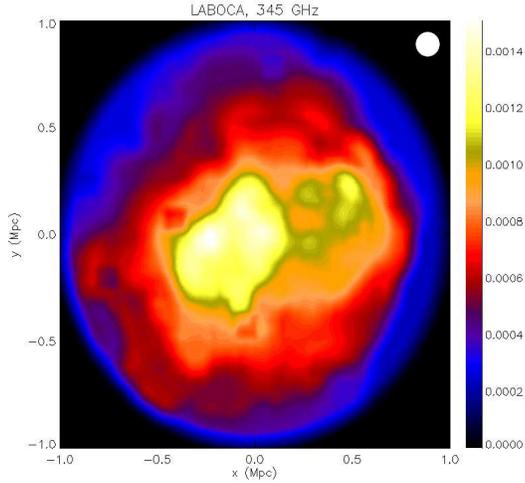}
\caption{The SZ intensity map $\Delta I/I_{0}$ for the cluster
1E0657-558 at 345 GHz smoothed to the resolution of LABOCA of
19$^{\prime\prime}$} \label{F3}
\end{figure}

\begin{figure}
\centering
\includegraphics[angle=0, width=7.0cm]{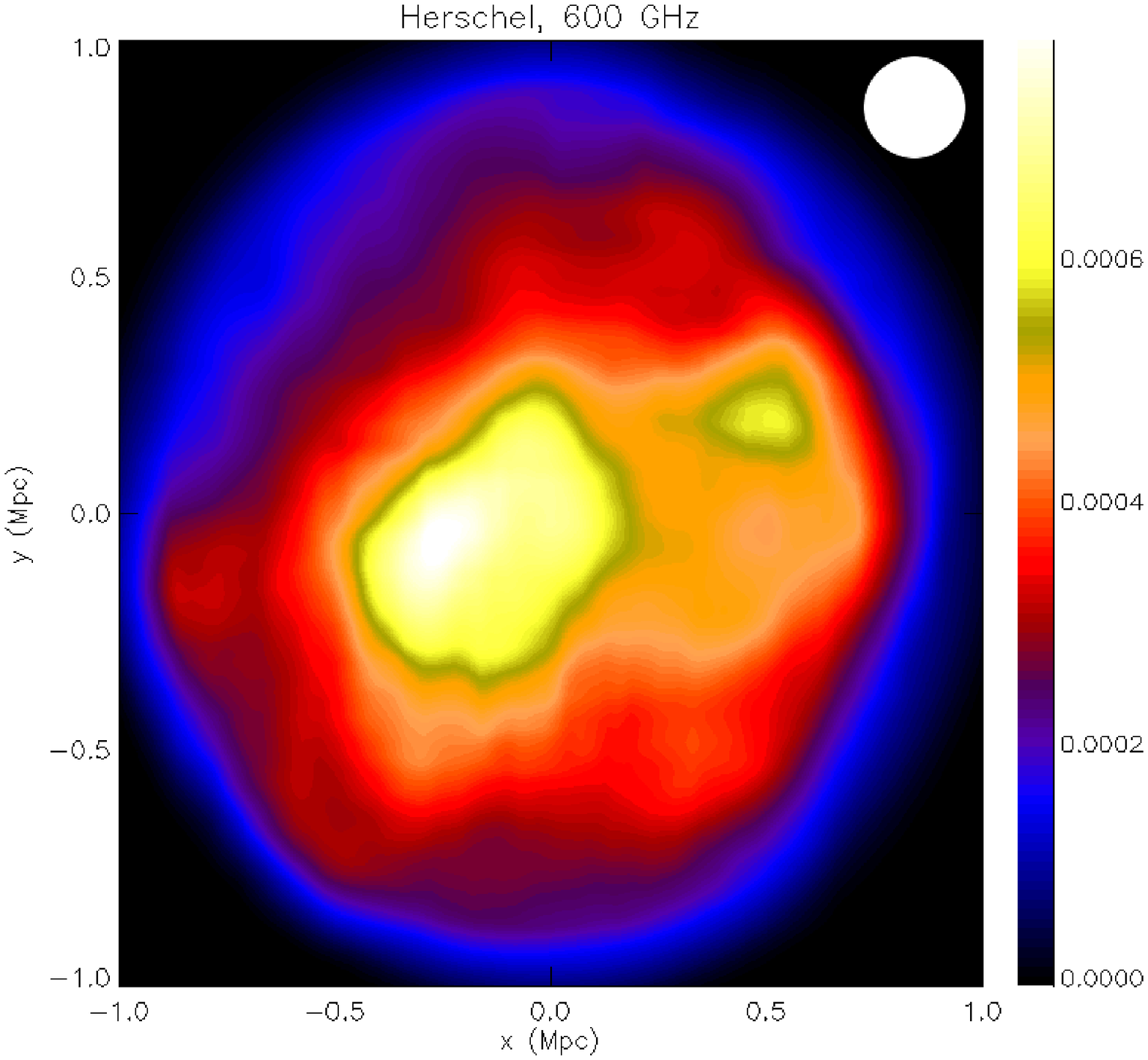}
\caption{The SZ intensity map $\Delta I/I_{0}$ for the cluster
1E0657-558 at 600 GHz smoothed to the resolution of HERSCHEL-SPIRE
of 36$^{\prime\prime}$} \label{F4}
\end{figure}

\begin{figure}
\centering
\includegraphics[angle=0, width=7.0cm]{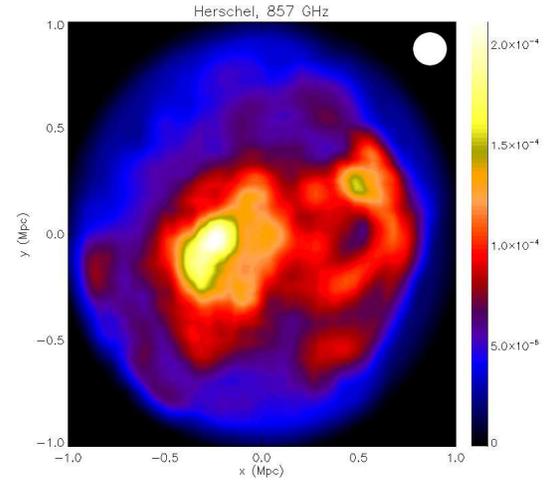}
\caption{The SZ intensity map $\Delta I/I_{0}$ for the cluster
1E0657-558 at 857 GHz smoothed to the resolution of HERSCHEL-SPIRE
of 25$^{\prime\prime}$}\label{F44}
\end{figure}

The cluster 1E0657-558 has been recently observed by the APEX-SZ
instrument at a frequency of 150 GHz (Halverson et al. 2009). The
observed SZ intensity map at 150 GHz has an effective resolution of
85$^{\prime\prime}$ FWHM and is shown in Fig. 3 of Halverson et al.
(2009). This intensity map does not show any significant SZ
intensity change in the western subcluster region compared to that
of the eastern subcluster.

To produce the SZ intensity maps at frequencies of 345 GHz, 600 GHz,
and 857 GHz we use the 2D temperature and pressure maps found by
Chandra (see Fig. 4 of Million \& Allen 2009). Since the X-ray data
are projected spectra, temperature and pressure information
corresponds to a column of gas through the cluster. We use the
average MEKAL normalization K \footnote{$K=\frac{10^{-14}}{4\pi
D^2_\mathrm{A} (1+z)^2}\int n_{\mathrm{e}} n_{\mathrm{H}}dV$, where
$D_{\mathrm{A}}$ is the angular diameter distance, $z$ is the
redshift, $n_{\mathrm{e}}$ and $n_{\mathrm{H}}$ are electron and
hydrogen electron densities} to derive the pressure in the energy
density units from the value of $k_{\mathrm{b}}T\times \sqrt{K/A}$
given in Fig. 4 of Million \& Allen (2010), where A is the area of
the region. Applying the operation of translation we produce 3D
density and temperature maps. Using the Wright formalism and the
approach described in Sect. 2, we calculate the SZ intensity maps
and smooth the SZ intensity maps at frequencies of 345 GHz, 600 GHz,
and 857 GHz to obtain the resolution of LABOCA and HERSCHEL-SPIRE,
respectively. The SZ intensity maps at 345 GHz, 600 GHz, 857 GHz of
the cluster 1E0657-558 smoothed to the resolution of
19$^{\prime\prime}$, 36$^{\prime\prime}$, and 25$^{\prime\prime}$
(FWHM), respectively, are shown in Figs. \ref{F3}, \ref{F4} and
\ref{F44}.

A comparison of Figs. \ref{F3} and \ref{F44} suggests that the
spatial feature on the SZ intensity map, caused by the presence of
the cool western subcluster, is seen on the SZ intensity map at a
frequency of 857 GHz, while this feature does not appear on the SZ
intensity map at a frequency of 345 GHz. This is because the
relativistic SZ corrections are small for low temperature plasmas
compared to those for high temperature plasmas. Since the
relativistic SZ corrections are the dominant contribution to the SZ
effect at very high frequencies for high-temperature plasma, the
cool `Bullet' substructure is seen as a decrement in the SZ
intensity map at 857 GHz. Figure \ref{F4} shows that the SZ
intensity at 600 GHz from the cool western subcluster has a local
decrement caused by the cool `Bullet' substructure. We calculate the
ratio of the SZ intensities at 600 GHz to 345 GHz in Sect. 3.4 and
find that the ratio is the lowest at the location of the cool
`Bullet' substructure. A comparison of Figs. \ref{F3} and \ref{F44}
also shows the presence of a bright, hot region in the eastern
subcluster in Fig. \ref{F44}. This hot region is bright at 857 GHz
because the relativistic SZ corrections are the highest in the
hottest regions and increase with frequency.

We find that the combined analysis of the LABOCA and HERSCHEL-SPIRE
SZ intensity maps is a method to reveal the cool `Bullet'
substructure. The difference between the calculated SZ signal at 857
GHz at the location of the cool `Bullet' substructure and the
expected SZ increment at 857 GHz at this location, derived from the
SZ intensity map at 345 GHz by using the Kompaneets approximation,
is $\Delta I/I_{0}\approx 5.0\times10^{-5}$. This difference
corresponds to 0.19 mJy/beam. The SZ intensity map observable by
HERSCHEL-SPIRE at 857 GHz should be more consistent with the SZ
intensity map shown in Fig.\ref{F44} than with that obtained by
APEX-SZ at 150 GHz. Using the SZ intensity maps at 345 GHz and 600
GHz, we find that the difference between the calculated and expected
(derived from the SZ intensity map at 345 GHz by using the
Kompaneets approximation) SZ signals at 600 GHz in the region of the
cool western subcluster is $\Delta I/I_{0}\approx 4.0\times 10^{-5}$
and corresponds to 0.35 mJy/beam. Foreground-free SZ measurements by
HERSCHEL-SPIRE at 600 GHz with an on-source integration time of 10
hrs are required to reduce the instrumental noise for a detection of
the difference in the SZ map morphologies caused by the presence of
the cool `Bullet' substructure. We also find that the bright, hot
region (in the eastern subcluster) on the foreground-free SZ
intensity map at a frequency of 857 GHz can be unveiled by
HERSCHEL-SPIRE with an on-source integration time of 2 hrs.

\subsection{The morphological SZ effect from the cluster Abell 2219 }

We calculate the SZ intensity maps in the Wright formalism for the
cluster A2219 to show that the morphological SZ effect from this
cluster could be revealed by observations of the SZ signal at
frequencies of 345 GHz, 600 GHz, and 857 GHz.

The massive galaxy cluster A2219 at redshift z=0.22 is an
interesting target to observe the SZ effect at very high
frequencies, since the average temperature of this cluster is very
high (Million \& Allen 2009). The temperature map of A2219 (see Fig.
A1 from Million \& Allen 2009) shows two extremely hot regions. The
first is associated with the merger shock front, and the second is
that coincides with the X-ray surface brightness peak. These
features are likely associated with recent merger activity.
Therefore, SZ relativistic corrections should be important in these
regions and these high temperature regions should be bright on the
SZ intensity maps at very high frequencies. The angular extent of
the cluster A2219 observed by Chandra is close to
$6^{\prime}\times6^{\prime}$ and can be therefore covered by the
FOVs of LABOCA and HERSCHEL-SPIRE instruments.

\begin{figure}
\centering
\includegraphics[angle=0, width=7.0cm]{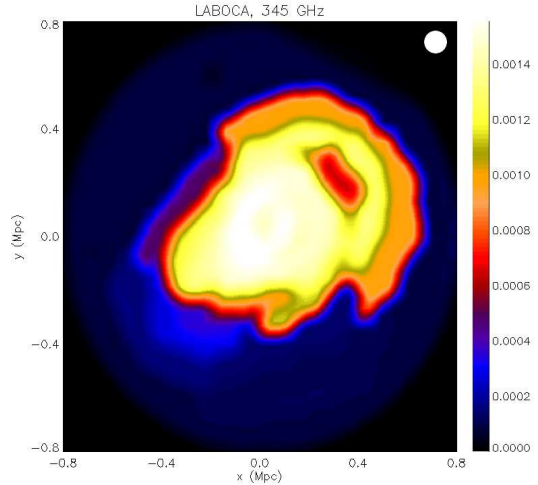}
\caption{The SZ intensity map $\Delta I/I_{0}$ for the A2219 cluster
at a frequency of 345 GHz smoothed to the resolution of LABOCA of
19$^{\prime\prime}$} \label{F5}
\end{figure}

\begin{figure}
\centering
\includegraphics[angle=0, width=7.0cm]{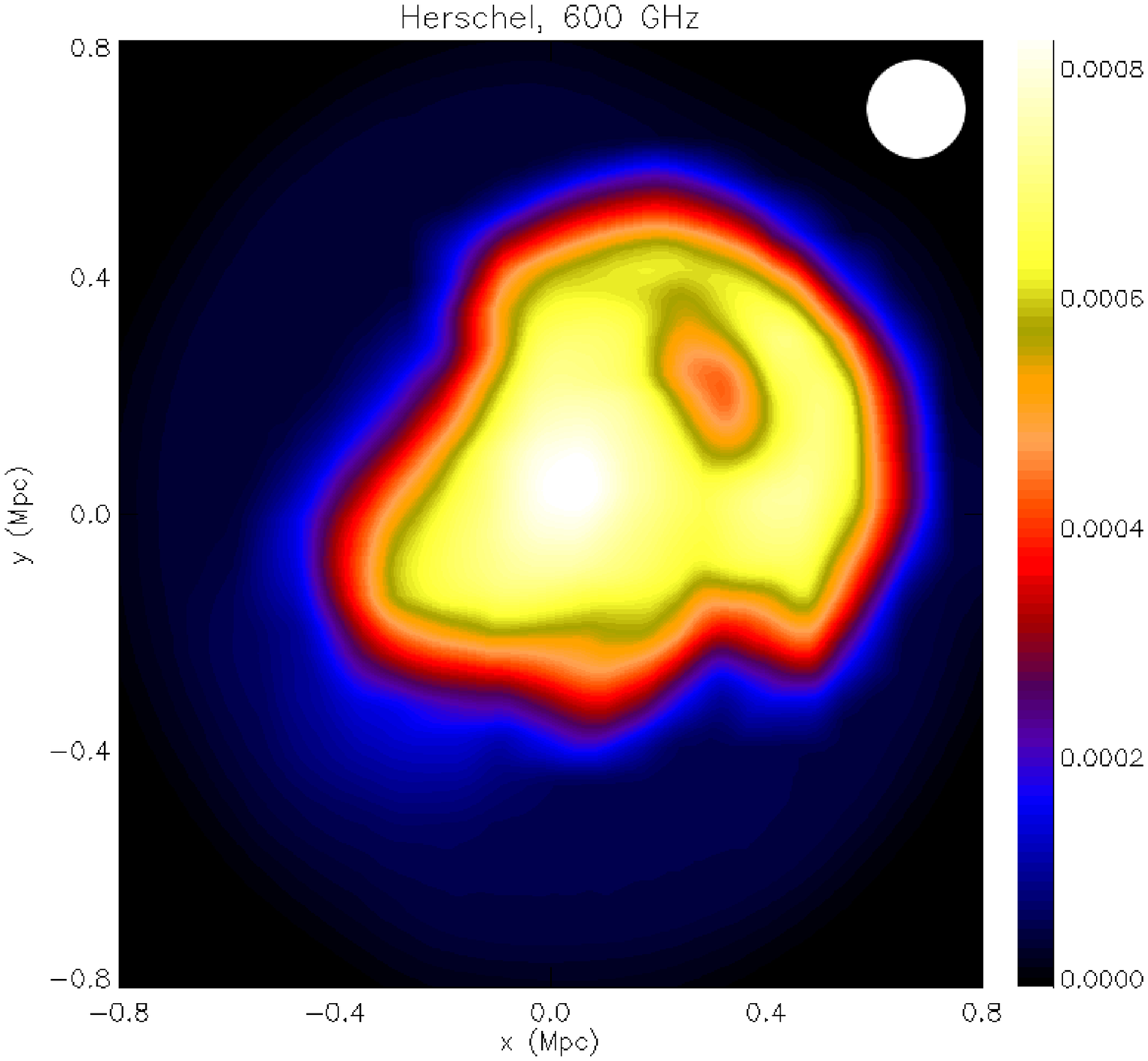}
\caption{The SZ intensity map $\Delta I/I_{0}$ for the A2219 cluster
at a frequency of 600 GHz smoothed to the resolution of
HERSCHEL-SPIRE of 36$^{\prime\prime}$} \label{F6}
\end{figure}

\begin{figure}
\centering
\includegraphics[angle=0, width=7.0cm]{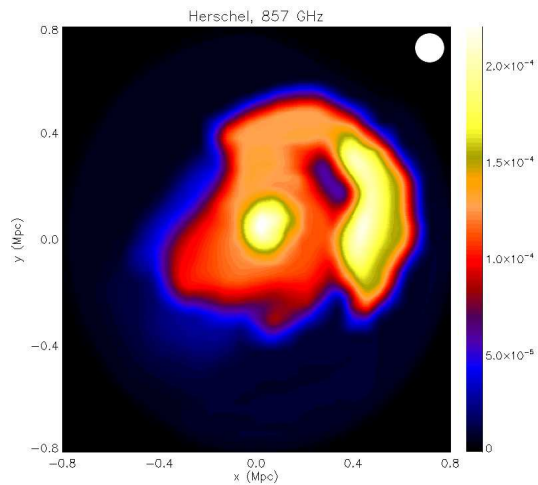}
\caption{The SZ intensity map $\Delta I/I_{0}$ for the A2219 cluster
at a frequency of 857 GHz smoothed to the resolution of
HERSCHEL-SPIRE of 25$^{\prime\prime}$} \label{F66}
\end{figure}

To produce the SZ intensity maps at frequencies of 345 GHz, 600 GHz,
and 857 GHz we use the 2D temperature and pressure maps found by
Chandra (from Million \& Allen 2009) and consider them as projected.
We construct 3D density and temperature maps of A2219 by analogy
with those for the cluster 1E0657-558 discussed above. Using the
Wright formalism and the approach described in Sect. 2, we calculate
the SZ intensity maps and smooth the SZ intensity maps at
frequencies of 345 GHz, 600 GHz, and 857 GHz to obtain the
resolution of LABOCA and HERSCHEL-SPIRE, respectively. The SZ
intensity maps at frequencies of 345 GHz, 600 GHz, and 857 GHz of
the cluster A2219 smoothed to the resolution of 19$^{\prime\prime}$,
36$^{\prime\prime}$ and 25$^{\prime\prime}$ (FWHM), respectively,
are shown in Figs. \ref{F5}, \ref{F6} and \ref{F66}.

Comparing Fig.\ref{F5} with Fig.\ref{F66}, we find that the SZ
intensity map at a frequency of 345 GHz is more uniform than that at
frequency of 857 GHz. The SZ intensity map at a frequency of 857 GHz
shows evidence for a large-scale shock heated region and a very hot
region coincident with the peak of the X-ray surface brightness
found by Million \& Allen (2010). Therefore, observations of the SZ
effect at a frequency of 857 GHz should confirm the presence of
these hot plasma regions in A2219. High-frequency analysis of the SZ
effect morphology will permit us to discover exceptionally hot gas
regions in massive galaxy clusters. A comparison of the SZ intensity
maps at 345 GHz, 600 GHz, and 857 GHz shows that the brightness of
the shock region on the SZ intensity maps increases with frequency
(we will perform a comparison of the SZ maps of the cluster A2219 at
frequencies of 345 GHz and 600 GHz by means of the SZ intensity
ratio in Sect. 3.4). We find that the the on-source integration time
of 2 hrs of HERSCHEL-SPIRE at a frequency of 857 GHz is required to
reduce the instrumental noise for unveiling both the shock and hot
cental region in the galaxy cluster A2219.

We conclude that the galaxy cluster A2219 is an interesting target
to analyze by means of the SZ effect at very high frequencies and
exhibits regions of very hot gas which could potentially be observed
by SZ observations at frequencies of 600 GHz and 857 GHz. Therefore,
an analysis of massive galaxy clusters by means of the SZ effect at
very high frequencies is a promising tool to study gas-dynamics
processes.

\subsection{The ratio of the SZ intensity at frequency 600 GHz to that at frequency 345 GHz}

Studying of the ratio of the SZ intensities at two frequencies is a
way to measure the gas temperature (see Prokhorov et al. 2010b). We
checked that the ratio of the SZ intensity at a frequency of 600 GHz
to that at a frequency of 345 GHz is a monotonically increasing
function of gas temperature and, therefore, measurements of this
ratio allows us to unambiguously determine the gas temperature. In
this section, we demonstrate that the SZ intensity ratio can be used
to quantify the morphological SZ effect.

In the framework of the Kompaneets approximation, the SZ intensity
ratio at any two given frequencies does not depend on gas
temperature. Therefore, in this approximation the SZ intensity map
at a dimensionless frequency of $x_{2}$ can be derived from the SZ
intensity map at a dimensionless frequency of $x_{1}$ by using the
rule $\Delta I_{2}=\Delta I_{1} g(x_{2})/g(x_1)$ and SZ intensity
maps look similarly for any frequencies. SZ relativistic corrections
contribute to the SZ signal and the morphology of SZ intensity maps
at two different frequencies are different in the relativistically
correct formalism. Thus, we propose to use the SZ intensity ratio as
a quantity which characterizes the morphological SZ effect. Note
that the morphological SZ effect is caused by SZ relativistic
corrections.

\begin{figure}
\centering
\includegraphics[angle=0, width=6.8cm]{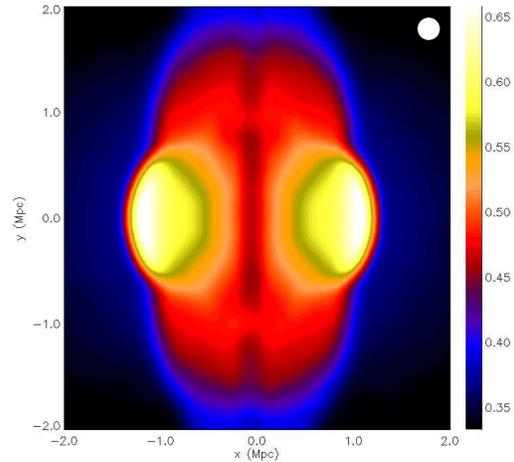}
\caption{The ratio of the SZ intensity at frequency 600 GHz to that
at frequency 345 GHz for the simulated galaxy cluster.} \label{F7}
\end{figure}

\begin{figure}
\centering
\includegraphics[angle=0, width=6.8cm]{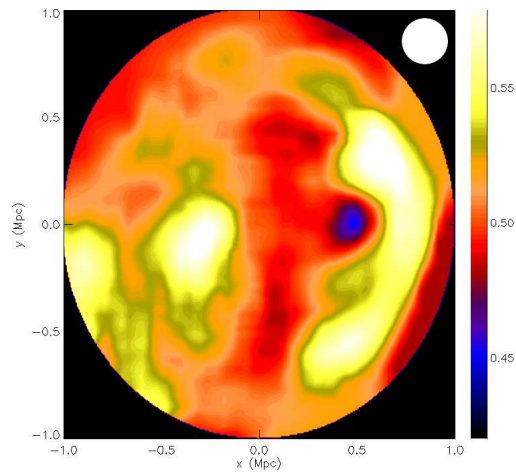}
\caption{The ratio of the SZ intensity at frequency 600 GHz to that
at frequency 345 GHz for the galaxy cluster 1E0657-558.} \label{F8}
\end{figure}

\begin{figure}
\centering
\includegraphics[angle=0, width=6.8cm]{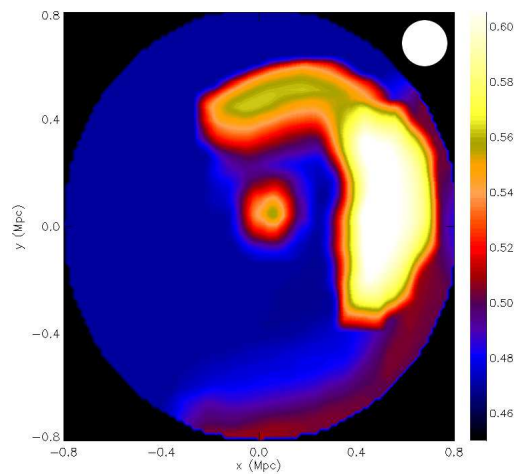}
\caption{The ratio of the SZ intensity at frequency 600 GHz to that
at frequency 345 GHz for the galaxy cluster A2219} \label{F9}
\end{figure}

Using the SZ intensity maps at frequencies of 345 GHz and 600 GHz
derived in the framework of the Wright formalism in Sects. 3.1, 3.2,
and 3.3, we calculate the SZ intensity ratio at frequencies 600 GHz
to 345 GHz for the galaxy clusters studied above. The derived maps
of the SZ intensity ratio are shown in Figs. \ref{F7}, \ref{F8}, and
\ref{F9} for the simulated galaxy cluster, 1E0657-558, and A2219,
respectively. These maps are smoothed to the resolution of
HERSCHEL-SPIRE of 36$^{\prime\prime}$ at a frequency of 600 GHz,
since this angular resolution is lower than that of LABOCA at a
frequency of 345 GHz.

Figure \ref{F7} shows the presence of two regions with high values
of the SZ intensity ratio in the simulated galaxy cluster. These
regions correspond to the hottest regions of the simulated cluster
(see Akahori \& Yoshikawa 2009) and these regions on the SZ
intensity map at a frequency of 600 GHz (see Fig. \ref{F2}) are the
most brightest. Therefore, the SZ intensity maps at frequencies of
345 GHz and 600 GHz look different. The map of the SZ intensity
ratio for the galaxy cluster 1E0657-558 which is shown in Fig.
\ref{F8} has a region with low values of the SZ intensity ratio and
this region coincides with the position of the cold substructure
`Bullet'. Therefore, the cold substructure `Bullet' is clearly seen
on the SZ intensity maps at high frequencies (600 GHz and 857 GHz)
as a decrement of the SZ intensity. Note that Fig. \ref{F8} shows
evidence for a high-temperature region in the eastern subcluster and
a merger shock. Figure \ref{F9} shows that the SZ intensity ratio
for the galaxy cluster A2219 is high at the position of the shock
wave and in the central region of the map. This is because the gas
temperature is higher in these regions compared with the average gas
temperature in the galaxy cluster A2219. Thus, the SZ intensity in
these regions are bright at high frequencies (600 GHz and 857 GHz)
compared with the average SZ intensity.

We conclude that the maps of the SZ intensity ratio at frequencies
600 GHz to 345 GHz demonstrate that the morphologies of the SZ
intensity maps at these frequencies are different. This permits us
to quantify the SZ morphological effect.

\subsection{The generalized Kompaneets formalism}

In this section, we consider an alternative method to calculate the
SZ effect including SZ relativistic corrections which was proposed
by Challinor \& Lasenby (1998) and Itoh et al. (1998). This method
is based on a study of the generalized Kompaneets equation. An
analytical expression for the SZ effect including relativistic terms
up to O($\Theta^5_{\rm{e}}$) terms was derived by Itoh et al.
(1998), where $\Theta_{\rm{e}}=k_{\rm{b}}T_{\rm{e}}/m_{\rm{e}}c^2$
is the relativistic expansion parameter.

To compare the SZ effect derived from the Wright formalism with that
derived from the generalized Kompaneets formalism, we use Eqs.
(2.25-2.30) of Itoh et al. (1998). Note that the terms describing SZ
relativistic corrections up to O($\Theta^3_{\rm{e}}$) given by Eqs.
(2.26-2.28) of Itoh et al. (1998) agree with the results obtained by
Challinor \& Lasenby (1998), the terms describing SZ corrections
correct to higher order in the expansion parameter $\Theta_{\rm{e}}$
are given only by Itoh et al. (1998).

The SZ effect in the framework of the generalized Kompaneets
formalism can be written in the form given by Eq. (\ref{form}),
where the relativistic spectral function $G(x, T_{\mathrm{e}})$
equals
\begin{equation}
G(x, T_{\mathrm{e}})=\frac{x^4\exp(x)}{(\exp(x)-1)^2}\sum_{n\geq0}
Y_{n}\Theta^n_{\rm{e}},
\end{equation}
where the spectral functions of $Y_0$, $Y_1$, $Y_2$, $Y_3$, and
$Y_4$ are given by Eqs. (2.26-2.30) of Itoh et al. (1998).

\begin{figure}
\centering
\includegraphics[angle=0, width=7.0cm]{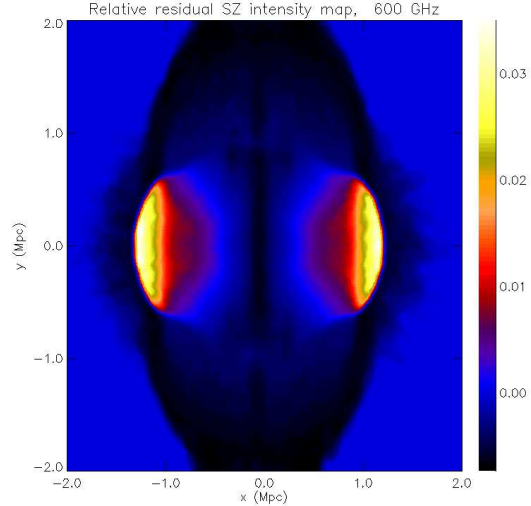}
\caption{Relative residual SZ intensity map at a frequency of 600
GHz for the simulated cluster between the SZ signals derived from
the Wright and extended Kompaneets (correct to third order in
$k_{\rm{b}}T_{\rm{e}}/(m_{\rm{e}} c^2)$) formalisms.} \label{F10}
\end{figure}

To demonstrate that the generalized Kompaneets formalism can also be
used to derive the SZ intensity maps, we calculate the SZ effect at
a frequency of 600 GHz for the simulated galaxy cluster by using
both the Wright and generalized Kompaneets formalisms. We estimate
the number of terms in the expansion of $\sum_{n\geq0}
Y_{n}\Theta^n_{\rm{e}}$ in the framework of the generalized
Kompaneets formalism which are necessary to be taken into account
for calculating the SZ effect with a precision better than 5\%. We
find that the four terms in the expansion of $\sum^{3}_{n=0}
Y_{n}\Theta^n_{\rm{e}}$ are sufficient to calculate the SZ effect
with a precision better than 5\% for the simulated cluster. We
calculate the relative residual SZ signal at a frequency of 600 GHz
from the simulated cluster, which is determined by the expression
$(\Delta I_{W}-\Delta I_{gK})/\Delta I_{W}$, where $\Delta I_{W}$ is
the SZ intensity derived by using the Wright formalism and $\Delta
I_{gK}$ is the SZ intensity derived by using the generalized
Kompaneets formalism correct to third order in
$k_{\rm{b}}T_{\rm{e}}/(m_{\rm{e}} c^2)$. The relative residual SZ
intensity map at a frequency of 600 GHz for the simulated cluster
between the SZ signals derived in the frameworks of the Wright and
generalized Kompaneets (correct to third order in
$k_{\rm{b}}T_{\rm{e}}/(m_{\rm{e}} c^2)$) formalisms is shown in Fig.
\ref{F10}. Figure \ref{F10} shows that the precision of calculation
of the SZ effect by means of the generalized Kompaneets formalism
(correct to the third term in $k_{\rm{b}}T_{\rm{e}}/(m_{\rm{e}}
c^2)$) is better than 5\% for the simulated cluster. The presence of
positive and negative regions on the residual SZ intensity map are
caused by the fact that the frequency of 600 GHz is close to the
frequency at which the numerical (precisely calculated) and
analytical (approximately calculated) curves shown in Fig. 5 of Itoh
et al. (1998) are crossing.

We calculate the residual SZ intensity map at a frequency of 600 GHz
for the simulated cluster between the SZ signals derived from the
Wright and (non-extended) Kompaneets formalisms and find that the
difference in the SZ intensities is around 30\% in the regions where
the difference in the SZ intensities in Fig. \ref{F10} is maximal.
Since the precision of the generalized Kompaneets formalism is
sufficiently high for a plasma with temperature of 20 keV to
estimate the relativistic SZ corrections, we conclude that the
generalized Kompaneets formalism can be used instead of the Wright
formalism to derive the SZ intensity maps.

\section{Conclusions}

Merging galaxy clusters are an interesting astrophysical laboratory
for studying gas-dynamics processes. Mergers of galaxy clusters are
very energetic astrophysical events in which huge gravitational
energy is released. In the course of a merger, a significant portion
of this energy is dissipated by merger shock waves. This leads to a
heating of the gas to higher temperatures. On the other hand, cool,
dense gas clouds in the ambient high temperature gas of clusters,
such as the `Bullet' in the cluster 1E0657-558, are also present in
merging clusters. We show that observations of the SZ effect at very
high frequencies can provide a method to study these phenomena by
analyzing the relativistic corrections to the SZ effect.

To produce a realistic high temperature merging cluster, we have
used the cosmological simulation from Akahori \& Yoshikawa (2010).
We have found that the simulated cluster undergoes a violent merger,
which drives the cluster gas to higher temperatures. We have
incorporated the relativistic Wright formalism for modeling the SZ
effect in the numerical simulation using the procedure proposed by
Prokhorov et al. (2010a). We have calculated the SZ intensity maps
at frequencies of 345 GHz, 600 GHz, and 857 GHz at which the SZ
effect is observable by LABOCA and HERSCHEL-SPIRE, respectively. We
have demonstrated that the morphology of the SZ intensity maps at
these frequencies are different (i.e., there is evidence for a
morphological SZ effect), since SZ relativistic corrections are
large for a high temperature plasma compared with those for a low
temperature plasma and are the dominant contribution to the SZ
effect at very high frequencies. The morphological SZ effect from
the simulated galaxy cluster is caused by a violent merger activity.\\
Using the Wright formalism and the density and temperature maps of
the cluster 1E0657-558 obtained by Chandra (Million \& Allen 2009),
we calculated the SZ intensity maps of the cluster 1E0657-558 at
frequencies of 345 GHz, 600 GHz, and 857 GHz. We found that the SZ
intensity map at 857 GHz has a spatial feature caused by the
presence of the cold substructure `Bullet' seen in the X-ray surface
brightness map. However, this cold substructure is not present on
the SZ intensity map at 345 GHz. This is a consequence of the
relativistic corrections of the SZ effect and shows that
observations of the SZ intensity maps at very high frequencies are
promising to reveal the complex substructures within massive galaxy
clusters.

We have also calculated the SZ effect from the cluster A2219 at
frequencies of 345 GHz, 600 GHz, and 857 GHz using the Wright
formalism and the density and temperature maps obtained by Chandra
(Million \& Allen 2009). The SZ intensity maps at high frequencies
show evidence for a large-scale shock heated region and a very hot
region coincident with the peak of the X-ray surface brightness.
Therefore, an analysis of the SZ signal at high frequencies is a
promising method for unveiling high temperature regions in massive
merging clusters.

This paper is a theoretical study of the morphological changes in
the images of the SZ effect in bright merging galaxy clusters caused
by the contribution of the relativistic corrections to the SZ effect
and by variations in the temperature of the intracluster medium. We
show that integration times of several hours are required to reveal
the differences in the SZ map morphologies of the considered
clusters. Note that foreground-free maps predicted for
HERSCHEL-SPIRE are not a fully realistic realization. These maps
should be clean of contaminating sources up to the level that the
structure remaining in the maps is predominantly the SZ effect
signal. The remaining unresolved sources are going to be both a
source of uncertainty and a bias for the measurements.
The differences in the SZ intensity maps at 345 GHz, 600 GHz, and
857 GHz can be marginally revealed by the combined analysis of the
LABOCA and HERSCHEL-SPIRE intensity maps. We conclude that the SZ
intensity maps at very high frequencies will permit us to reveal
very hot plasma regions in galaxy clusters.

The energy ranges of the modern X-ray instruments of (0.5 keV -- 10
keV) constrain our ability to analyze the most massive galaxy
clusters with temperatures of ($k_{\mathrm{b}} T_{\mathrm{e}}\approx
15$ keV) and the hottest regions in these clusters. We suggest that
the morphological SZ effect will provide us with an important
additional tool to study the merger activity in the most massive
galaxy clusters.

\section*{Acknowledgments}
We are grateful to Steven Allen for valuable suggestions and
discussions and thank the referee for valuable suggestions.

\end{document}